\documentclass{aa}

\usepackage{float,graphics,psfig,epsf,rotate}

\def\gx{GX~339--4}
\def\cx{Cyg~X--1}
\def\grs{GRS~1915+105}

\def\1e{1E~1740.7--2942}

\def\xd{$\times$ 10}
\def\batse{{\em BATSE}}

\begin{document}

   \thesaurus{     
              (08.09.2 \gx;  
               13.07.2; 
               13.18.5; 
               13.25.5;
               02.01.2; 
               02.02.1)} 

   \title{Coupling of the X-ray and radio emission in the black hole candidate and compact jet source \gx}

   \author{S. Corbel\inst{1}, R.P. Fender \inst{2}, A.K. Tzioumis \inst{3}, 
	M. Nowak \inst{4}, V. McIntyre \inst{5}, 
	P. Durouchoux \inst{1} \and R. Sood \inst{6}}
   
   \offprints{corbel@discovery.saclay.cea.fr}

   \institute{
	CEA-Saclay, Service d'Astrophysique, F-91191 Gif sur Yvette Cedex, France
\and
	Astronomical Institute `Anton Pannekoek', University of Amsterdam, and Center for High Energy Astrophysics, Kruislaan 403, 1098 SJ Amsterdam, The Netherlands
\and
        Australia Telescope National Facility, CSIRO, Paul Wild Observatory, Narrabri NSW 2390, Australia
\and
        JILA, University of Colorado, Campus Box 440, Boulder, CO 80309-0440, USA
\and
        Department of Astrophysics, School of Physics, University of Sydney, NSW 2006, Australia
\and
	School of Physics, Australian Defence Force Academy, Canberra, ACT 2600, Australia
}

   \date{Received ????; accepted ????}

   \titlerunning{\gx}

   \authorrunning{Corbel et al.}

   \maketitle

   \begin{abstract}

We report the results of a long-term campaign of radio, soft- and
hard- X-ray observations of the galactic black hole candidate \gx.  In
the Low-Hard X-ray state the system displays a strong 3-way linear
correlation between soft- and hard-X-rays and radio emission, implying
a coupling between the Comptonising corona and a
radio-emitting compact jet. In this state the radio emission is
linearly polarised at a level of around 2\%, with an almost constant
polarisation angle, indicative of a favored axis in this system
probably related to the compact jet and/or black hole spin axis. In
the Off X-ray state the radio emission declines with the X-ray
emission to below detectable levels, suggesting that it is simply a
lower-luminosity version of the Low-Hard state. In the High-Soft state
both the hard-X-ray and radio emission are suppressed.  We also note
that the transitions from the Low-Hard state to the High-Soft state
(and the reverse) are possibly associated with discrete ejection(s) of
expanding relativistic plasma.

 \end{abstract}


\section{Introduction}

The X-ray source \gx\ is believed to harbor a black hole,
based on the similarity of its X-ray spectral and timing properties
with the black hole system \cx\ (e.g. Samimi et al. 1979; Maejima et
al. 1984; Ilovaisky et al. 1986; Makishima et al. 1986).
Black hole candidates are known to exhibit transitions between various
X-ray states, distinguished by their different spectral and timing
properties. Five distinct X-ray states have been reported, namely
the Off state, the Low-Hard state, the Intermediate state, the
High-Soft state and the Very High state. See Tanaka \& Lewin (1995)
for a review of these properties in the Low-Hard and High-Soft states.

\gx\ and GS~1124--683 (\cite{ebi94}) are the only X-ray sources
observed in all these states [see Miyamoto et al. (1991) for the Very
High state and M\'endez \& van der Klis (1997) for the Intermediate
state in \gx]. Two new X-ray transient sources have also
been observed in the Very High state, i.e XTE~J1748--288 (Revnivtsev et al. 2000)
and XTE~J1550--564 (e.g. Sobczak et al. 1999, Homan et al. 2000). The
superluminal sources GRO~J1655--40 and GRS~1915+105 may also display 
behaviour related to X-ray spectral state changes (e.g. M\'endez et al. 1998,
Belloni et al. 2000).
Spectral and timing properties of \gx\ in the Low-Hard
state can be found in Wilms et al. (1999) and Nowak et al. (1999).  In 
the Low-Hard state, a
peak in the low-frequency power spectrum is seen at a period of 240 days, and 
is possibly related to a precessing accretion disk (Nowak et al. 1999).

Black hole candidates can be divided into two different classes of
sources: the soft X-ray transients (SXTs) and the persistent black
hole candidates. SXTs (\cite{cha98}) are, by definition, transient and are usually associated
with a low mass and late type stellar companion. \gx\ is persistent in soft-X rays (as are
the other galactic black hole candidates \cx, 1E~1740.7--2942 and  GRS~1758--258), but at hard X-rays
\gx\ shows behaviour resembling that of the transient sources (\cite{har94,rub98}).
We note that LMC~X--1 and LMC~X--3 can also be considered as persistent black hole candidates 
in the Large Magellanic Cloud (\cite{tan95}).

Due to the faintness of the stellar companion relative to the
accretion disk, the spectral type of the stellar companion of \gx\ has
not been identified, but it is believed to be a low mass star with a
possible orbital period of 14.8 hours (\cite{cow87,cal92}). Soria et
al. (1999) suggested the true orbital period may be twice this value.
The distance to \gx\ is about 4 kpc (\cite{zdz98}) with an optical
extinction of 3.5 magnitudes (\cite{cor99}).  Optical emission in \gx\ also displayed
the state transitions (e.g. \cite{mot85}). \gx\ is usually bright and
variable (V $\approx$ 15-17 mag.) in the Low-Hard state, faint in the
Off state (V $\approx$ 20 mag.) and at an intermediate level in the
High-Soft state (V $\approx$ 16.5 mag.).  Most of the optical emission
is believed to be dominated by the accretion flow, but the physical
process behind its origin has not been fully understood (\cite{fab82,mot83}, Ilovaisky et al. 1986).

Although \gx\ has been studied extensively at high energies, little has been known until recently
about its properties in the radio regime. It was discovered as
a radio source in 1994 (Sood \& Campbell-Wilson 1994) and a possible jet like feature
has been reported by Fender et al. (1997a). In the Low-Hard state,
Hannikainen et al. (1998) found a correlation between the radio, soft
and hard X-rays emission on timescale of the order of 5 days over a
period of one year.  During the High-Soft state, we have shown that
the radio emission from \gx\ disappeared until the return to the
standard Low-Hard state (\cite{fen99b}).  This behaviour is
reminiscent of that observed in the 1996 High-Soft (or
Intermediate) state of \cx\ (Zhang et al. 1997).

This paper presents the results from a long term campaign of radio
observations of \gx\ with the Australia Telescope Compact Array and
the Molonglo Observatory Synthesis Telescope.  Following the
description of the characteristics of radio emission (light curve,
spectra, variability and polarisation) from \gx, we present evidence
for the existence of a compact jet in this system.  We then focus on
the behaviour of \gx\ in the various X-ray states, as observed in
radio, soft and hard X-rays.  We demonstrate a strong coupling in the
Low-Hard state of the compact jet with the Comptonising corona.

\section{Observations}

\subsection{Molonglo Observatory Synthesis Telescope}

The Molonglo Observatory Synthesis Telescope (MOST) consists of two
co-linear cylindrical paraboloid reflectors each 778 m long, 12 m wide, and
separated by 15 m (Robertson 1991).  It is aligned on an east-west
axis. The MOST operates at 843 MHz (35.6 cm) with a 3 MHz
bandwidth and synthesises a beam of 43$\times$43 cosec($\delta$)
arcsec$^2$ for a field of view at declination $\delta$. 
In a full 12-hours observation with a 23$\times$23 cosec($\delta$) arcmin$^2$
field of view, the rms in the final image is $\sim$ 0.6 mJy. 

The standard MOST pipeline (McIntyre \& Cram 2000) has been used for
calibration and imaging.  Following the procedure of Hannikainen et
al. (1998), we fitted the data from three sources besides \gx\ in each
observation and scaling the fluxes so that the sum of these three
reference sources remained constant (on the assumption that these
sources do not vary). The flux values in Table \ref{tab_radio} are the fitted peak
flux of \gx\ multiplied by a correction factor.  All point source fits
from the synthesized map were performed with the MIRIAD IMFIT routine.

Uncertainties in the flux densities are given by the rms variation
across the entire image, ignoring sources brighter than 10~mJy. This
somewhat overestimates the peak flux uncertainty for partial synthesis
($<12^h$), because there may be some large-scale variation across the
image due to incomplete cancellation of sidelobes.  We have set upper
limits to be 3 times this value.

All MOST results (excluding the ones presented in Hannikainen et
al. 1998) are summarized in Table \ref{tab_radio}.  MOST flux densities are plotted in
Figure \ref{fig_most} and in the top panels of Figures \ref{fig_a},
\ref{fig_b}, \ref{fig_c}, \ref{fig_d}, \ref{fig_e}, \ref{fig_f} and
\ref{fig_g}.
 
\subsection{Australia Telescope Compact Array}

Since 1994, we have been performing observations with the Australia
Telescope Compact Array (ATCA) a few times a year.  We also found,
using the ATCA archives, observations from 1991 and 1992. The ATCA
synthesis telescope is an east-west array consisting of six 22 m
antennas with baselines ranging from 31 m to 6 km.  The continuum
observations have been performed in two frequency bands (with a
total bandwidth of 128 MHz for 32 channels), usually at 4800 MHz (6.3
cm) and 8640 MHz (3.5 cm) simultaneously and with sporadic
observations at 2368 MHz (12.7 cm) and 1384 MHz (21.7 cm).  Various
array configurations have been used during these observations.

The ATCA has orthogonal linearly polarized feeds and full Stokes
parameters (I, Q, U, V) are recorded at each frequency.  B1934-638 was
used for absolute flux and bandpass calibration, while B1646--50 or
B1722--55 have been used as phase calibrators, in order to calibrate
the antenna gains and phases as a function of time and to determine
the polarisation leakages.  Data are normally integrated over 10 s
intervals.  \gx\ was systematically offset by 10$\arcsec$ from the
array phase center.  The editing, calibration, Fourier transformation,
deconvolution and image analysis were carried out with the
Multichannel Image Reconstruction, Image Analysis and Display (MIRIAD)
software package (\cite{sau95,sau98}), allowing multifrequency
synthesis of all four Stokes parameters.  All ATCA results are
summarized in Table \ref{tab_radio}. ATCA flux densities are plotted in Figure
\ref{fig_atca} in the top panels of Figures \ref{fig_a}, \ref{fig_b},
\ref{fig_c}, \ref{fig_d}, \ref{fig_e}, \ref{fig_f} and \ref{fig_g}.

  \begin{table*}
\centering
      \caption[]{Radio observations of \gx\ with ATCA and MOST. 
This table does not include the MOST measurements presented in Hannikainen et al. (1998) with the exception of 
observations simultaneous with ATCA. Upper limits are given at the three sigma level.}
         \label{tab_radio}
\begin{tabular}{clllllll}
            \hline
            \noalign{\smallskip}
                       &      & \multicolumn{4}{c}{ATCA} & MOST \\
            \noalign{\smallskip}
\cline{3-6}
\cline{7-7}
            \noalign{\smallskip}
            Date       &  MJD   & 8640 MHz       & 4800 MHz      & 2350 MHz & 1440 MHz & 843 MHz &  \\
             (UT Time) &  (days) & (mJy)         & (mJy)         & (mJy)    & (mJy)    & (mJy)   & spectral index \\
            \noalign{\smallskip}
            \hline
            \noalign{\smallskip}
1991:12:01    & 48591.50    & 1.10 $\pm$ 0.10  & 1.60 $\pm$ 0.12 &        -     &         -    &-       &--0.64  \\
1991:12:13    & 48603.50    & 1.90 $\pm$ 0.10  & 2.10 $\pm$ 0.10 &        -     &         -    &-       &--0.17\\
1992:04:21    & 48733.50    & 0.60 $\pm$ 0.08  & 0.40 $\pm$ 0.06 &        -     &         -    &-       &+0.69\\
1994:11:15    & 49671.50    & 11.8 $\pm$ 0.10  & 11.7 $\pm$ 0.10 &        -     &         -    &-       &+0.01\\
1995:09:03    & 49963.50    & 5.20 $\pm$ 0.20  & 5.30 $\pm$ 0.20 &        -     &         -    &-       &--0.03\\
1995:09:22    & 49982.50    & 4.5 $\pm$ 1.0    & 4.6 $\pm$ 1.0   &        -     &         -    & -      &--0.04\\
1996:05:25    & 50228.13   & $<$ 0.6          & $<$ 1.2         & $<$ 1.8      & $<$ 1.8       & 2.3 $\pm$ 0.7 & \\
1996:05:27    & 50230.13   & $<$ 1.5          & $<$ 1.5         & $<$ 3.0      & $<$ 3.0       &        -       & -\\
1996:05:28    & 50231.13   &  -               &       -         &         -    &    -          &   $<$ 3.0 & -\\
1996:07:10    & 50274.67   & 6.20 $\pm$ 0.10  & 5.30 $\pm$ 0.10 &         -    &    -         &    -  &        +0.27\\
1996:07:13    & 50277.63   & 7.00 $\pm$ 0.10  & 6.40 $\pm$ 0.10 &         -    &    -        & 6.5 $\pm$ 0.6 & + 0.05\\
1996:07:14    & 50278.54   &  -               &       -         & 6.0 $\pm$ 0.1  & 5.2 $\pm$ 0.2 &               & +0.26\\
1996:12:15    & 50432.50   & 2.0 $\pm$ 1.0    & 2.0 $\pm$ 1.0   &         -    &    -        &     -         & +0.00\\
1997:02:04    & 50483.42   & 9.10 $\pm$ 0.10  &       -         &         -    &    -        & 7.0 $\pm$ 0.6 & +0.11\\
1997:02:06    & 50485.47   &  -               &       -         &7.4 $\pm$ 0.1 & 6.8 $\pm$ 0.2 & 6.6 $\pm$ 0.7 & +0.15\\
1997:02:11    & 50490.42   & 8.20 $\pm$ 0.10  &    -            &    -         &     -       & 5.5 $\pm$ 0.7  & +0.12\\
1997:02:18    & 50497.46   & 8.70 $\pm$ 0.20  &    -            & 6.0 $\pm$ 0.2  & 5.4 $\pm$ 0.2 & 5.6 $\pm$ 0.7  & +0.23\\
1997:04:30    & 50569.00   & 6.10 $\pm$ 0.10  & 4.70 $\pm$ 0.10 &   -          &    -        &     -         & +0.44\\
1997:07:22    & 50651.71   &  -               &       -         &         -    &    -        & 4.2 $\pm$ 0.7 & -\\
1997:08:01    & 50661.76   &            -     &       -         &         -    &    -        & 5.8 $\pm$ 0.8 & -\\
1997:08:10    & 50670.65   &            -     &       -         &         -    &    -        & 4.5 $\pm$ 0.7 & -\\
1997:08:15    & 50675.78   &            -     &       -         &         -    &    -        & 3.1 $\pm$ 0.8 & -\\
1997:09:14    & 50705.56   &            -     &       -         &         -    &    -        & 4.9 $\pm$ 0.7 & -\\
1997:10:18    & 50739.47   &            -     &       -         &         -    &    -        & 6.5 $\pm$ 0.7 & -\\
1997:11:16    & 50768.39   &            -     &       -         &         -    &    -        & 5.3 $\pm$ 0.7 & -\\
1997:12:07    & 50789.33   &              -   &       -         &         -    &    -        & 4.9 $\pm$ 0.8 & -\\
1998:01:01    & 50814.26   &            -     &       -         &         -    &    -        & 4.0 $\pm$ 0.8 & -\\
1998:01:09    & 50822.50   &    -             &    $<$ 0.3      &      -       &       -     & $<$ 1.5       & -   \\
1998:01:10    & 50823.30   &            -     &       -         &         -    &    -        & $<$ 2.7       & -\\
1998:01:14    & 50828.29   & 2.30 $\pm$ 0.05  & 2.90 $\pm$ 0.06 &   -          &    -        &     -         & --0.40\\
1998:01:26    & 50839.19   &            -     &       -         &         -    &      -      & $<$ 2.1       & -\\
1998:01:27    & 50840.79   & 0.26 $\pm$ 0.04  & 0.32 $\pm$ 0.04 &   -          &    -        &     -         & --0.35\\
1998:01:31    & 50844.34   &            -     &       -         &         -    &      -      &      $<$ 2.7 & -\\
1998:02:07    & 50851.34   &            -     &       -         &         -    &      -      &      $<$ 2.7 & -\\
1998:02:22    & 50866.12   &            -     &       -         &         -    &      -      &      $<$ 2.1 & -\\
1998:04:03    & 50907.01   &            -     &       -         &         -    &      -      &      $<$ 1.8 & -\\
1998:04:04    & 50908.29   & $<$ 0.18         &    $<$ 0.18     &      -       &       -     &     -        &  -   \\
1998:04:18    & 50921.97   &            -     &       -         &         -    &   -         &      $<$ 2.1 & -\\
1998:07:04    & 50998.78   &            -     &       -         &         -    &   -         &      $<$ 2.1 & -\\
1998:08:29    & 51054.60   &            -     &       -         &         -    &   -         &      $<$ 2.7 & -\\
1998:11:12    & 51129.58   &    $<$ 0.18      &      $<$ 0.12   &      -       &       -     &     -         &  -  \\
1999:01:13    & 51191.50   &    -             &      -          &      -       &    $<$ 1.0  &     -         & - \\
1999:01:18    & 51196.48   &            -     &       -         &          -   &   -         &       $<$ 2.7 & -\\
1999:01:23    & 51201.58   &    $<$ 0.2       &      $<$ 0.2    &      -       &       -     &     -         & - \\
1999:02:13    & 51222.50   & 4.60 $\pm$ 0.08  & 6.34 $\pm$ 0.08 &   -          &    -        &      -        &  --0.54 \\
1999:02:20    & 51229.33   &            -     &       -         &          -   &   -         & 5.2 $\pm$ 0.9 & -\\
1999:02:21    & 51230.30   &            -     &       -         &          -   &   -         & 6.2 $\pm$ 1.0 & -\\
1999:02:25    & 51234.30   &            -     &       -         &          -   &   -         & 5.5 $\pm$ 0.9 & -\\
1999:03:03    & 51241.21   & 5.74 $\pm$ 0.06  & 6.07 $\pm$ 0.06 &   -          &    -        &     -         & --0.10 \\
1999:04:02    & 51271.13   & 5.10 $\pm$ 0.07  & 4.75 $\pm$ 0.06 &   -          &    -        &      -        & +0.12 \\
1999:04:22    & 51291.08   & 3.11 $\pm$ 0.04  & 2.18 $\pm$ 0.05 &   -          &    -        &      -        & +0.60 \\
1999:05:14    & 51313.17   & 1.44 $\pm$ 0.04  & 1.25 $\pm$ 0.05 &   -          & $<$1.0        & $<$2.0          & +0.12 \\
1999:06:25    & 51355.17   & 0.34 $\pm$ 0.04  & 0.14 $\pm$ 0.03 &          -    & -          &     -         &  - \\
1999:07:07    & 51367.13   & 0.15 $\pm$ 0.03  & $<$ 0.13        &          -   &  -          &     -         &  -        \\
1999:07:29    & 51389.03   &   $<$ 0.09       &     -           &          -   &  -          &     -         &  -      \\
1999:08:17    & 51407.79   & 0.27 $\pm$ 0.06  &     -           &          -   &  -          &     -         &  -      \\
1999:09:01    & 51422.86   & 0.35 $\pm$ 0.05  &     -           &          -   &  -          &     -         &  -      \\
            \noalign{\smallskip}
            \hline
\end{tabular}
   \end{table*}

\subsection{CGRO BATSE}

\begin{figure*}
\resizebox{\hsize}{11cm} {\includegraphics{9629.f1}}
\caption{Radio light curve of  \gx\ plotted as a function of MJD (Modified Julian date) for the MOST observations. 
Upper limits are at the 1 $\sigma$ level.}
\label{fig_most}
\resizebox{\hsize}{11cm} {\includegraphics{9629.f2}}
\caption{Same as Figure \ref{fig_most}, but for the ATCA observations.}
\label{fig_atca}
\end{figure*}

The hard X sources in the sky have been continuously monitored by the
BATSE instrument (Large Area Detectors) aboard the {\em Compton Gamma
- Ray Observatory} (CGRO) since its launch in 1991.  BATSE consists of
eight identical uncollimated NaI(Tl) scintillation detector modules
arranged on the corners of CGRO (Fishman et al. 1989). Any portion of
the sky is occulted by the Earth several times a day.  Using an Earth
occultation technique (\cite{har94}), we were able to produce a
detector count rate light curve. With the BATSE instrumental response,
these count rates are then fitted using an optically thin thermal
bremsstrahlung (OTTB) model at a fixed temperature kT = 60 keV (following the
procedure of Rubin et al. 1998).  This enables us to determine the daily
flux values in the 20 - 100 keV energy band with typical 3 $\sigma$
sensitivity of 75 mCrab.  
OTTB is not the only model able to fit the hard X-ray spectra of \gx\ 
(see Bouchet et al. 1993, Harmon et al. 1994, Grabelsky et al. 1995, 
Rubin et al. 1998, Trudolyubov et al. 1998, Smith et al. 1999a) but 
it is sufficiently good for the purpose of getting a daily flux measurement
of \gx\ in the hard X-ray band.
Sources with declination less (or more) than
--(+) 43 degrees (which is the case for GX 339-4) will not be occulted
for a few days every $\sim$ 52 days, due to spacecraft configuration,
and this is responsible for the gaps in the BATSE data. Data points
very close to this transition would have very high rms and have been
flagged out.  Data with nearby interfering sources have also been
flagged. These flux measurements are plotted in the middle panel of
Figures \ref{fig_a}, \ref{fig_b}, \ref{fig_c}, \ref{fig_d},
\ref{fig_e}, \ref{fig_f} and \ref{fig_g} (or bottom panel when no soft
X-ray data are  available).

\subsection{RXTE ASM}

The All Sky Monitor aboard the {\em Rossi X-ray Timing Explorer}
(RXTE) scans the X-ray sky in three energy bands (1.3--3 keV, 3--5 keV
and 5--12.2 keV) five to ten times a day with typical dwell duration
of $\sim$ 90 s.  It provides a unique database on the activity in the
soft X-ray sky.  ASM consists of three Scanning Shadow Cameras with a
field of view of 6$\degr \times$ 90$\degr$ and a spatial resolution of
3\arcmin\ $\times$ 15\arcmin. The detectors are Position Sensitive
Proportional Counters with a total effective area of 90 cm$^2$.  See
Levine et al. (1996) for a detailed discussion of the RXTE ASM.
The full energy ASM data are displayed in the lower panels of Figure
\ref{fig_a}, \ref{fig_b}, \ref{fig_c}, \ref{fig_d}, \ref{fig_e},
\ref{fig_f} and \ref{fig_g}.

\section{Radio light curves}

\begin{figure}[b!]
       \resizebox{\hsize}{!}{\includegraphics{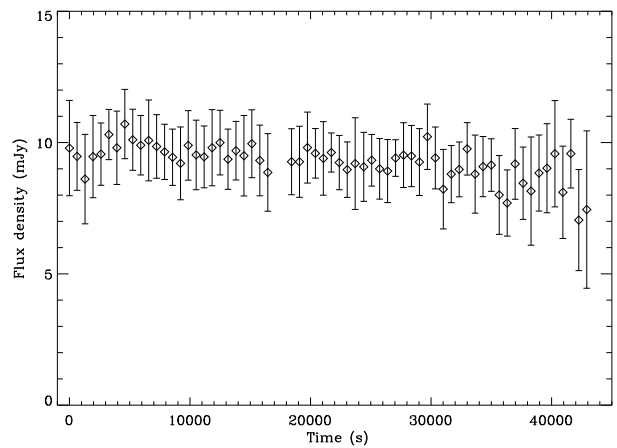}}
        \caption{Typical radio light curve of \gx\ at 9024 MHz during a 12 hours ATCA observations on 1997 February 18
        (MJD 50497). Each data point represents 10 minutes of integration time. Time is in seconds since the beginning
        of the observation.}
        \label{fig_lc1}
\end{figure}

\begin{figure*}[hbt]
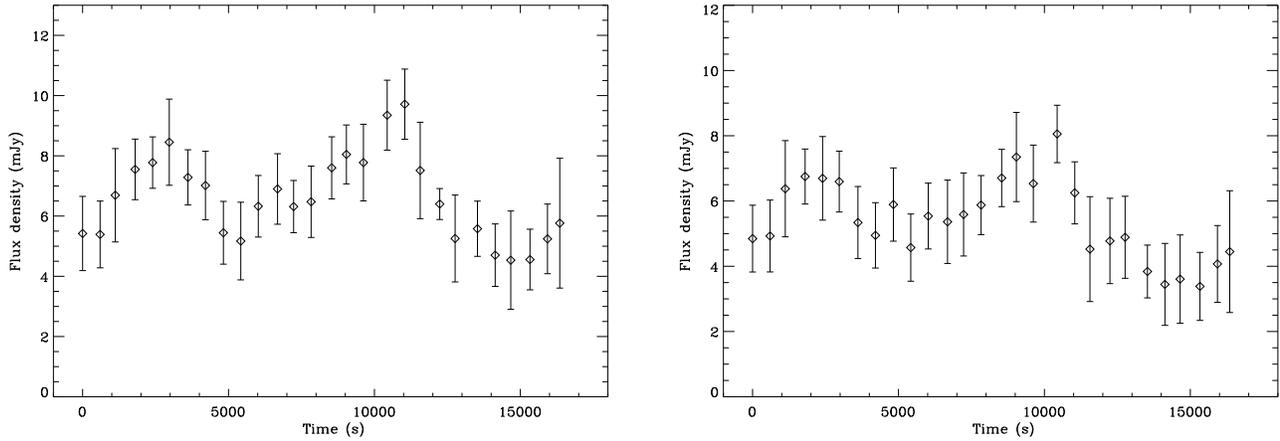

        \vspace{0cm}
        \hbox{\hspace{0cm}\psfig{figure=9629.f4a,width=8.8cm}\hspace{0cm}
                \psfig{figure=9629.f4b,width=8.8cm}}
        \caption{Radio light curves of \gx\ at 4800 MHz (left) and 8640 MHz (right) obtained by ATCA on 1999 February 13
        (MJD 51222).  Each data point represents 10 minutes of integration time. Time is in seconds since the beginning
        of the observation.}
        \label{fig_gx_lc2}
        \vspace{0cm}
\end{figure*}

\subsection{Long term variability}

Since the discovery of the radio counterpart by Sood \&
Campbell-Wilson (1994), \gx\ has been occasionally monitored by ATCA
and MOST.  
The radio light curves of \gx\ using the MOST telescope at 843 MHz and
the ATCA interferometer at 1384, 2368, 4800 and 8640 MHz are displayed 
in Figures \ref{fig_most} and \ref{fig_atca}. All these
observations have been performed when \gx\ was in the Off state,
Low-Hard state and High-Soft state.  \gx\ is detected at a mean level
of $\sim$ 5 mJy, with flux variations of $\sim$ 20 \% on a timescale of
days. Larger flux variations ($\sim$ 50 \%) occur on timescale of
weeks. The maximum flux density reached by \gx\ during our
observations (over a period of 8 years) is 12 mJy on 1994
November 15 at 4800 and 8640 MHz.

\begin{figure}[h!]
       \resizebox{\hsize}{!}{\includegraphics{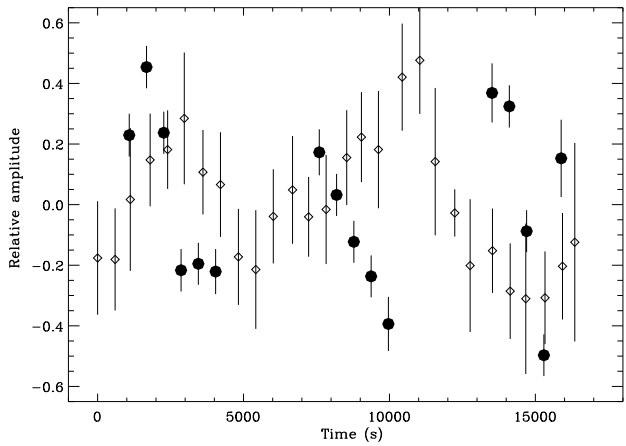}}
        \caption{Simultaneous X-ray (PCA - filled circle) and radio (4800 MHz - open square)
mean-subtracted light curves of \gx\ on 1999, February 13. They have further been divided by the 
mean and the X-ray light curve has also been multiplied by 30.
        }
        \label{fig_osc}
\end{figure}

\begin{figure}[h!]
       \resizebox{\hsize}{!}{\includegraphics{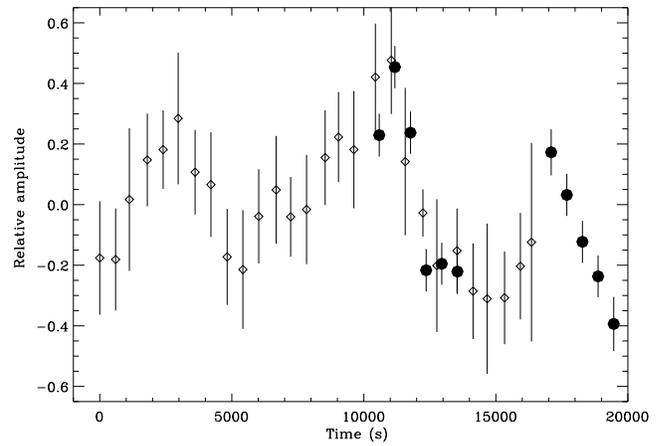}}
        \caption{Same as Figure \ref{fig_osc}, but with the X-ray lightcurve shifted by $\sim$ 10,000 seconds.
        }
        \label{fig_osc2}
\end{figure}

We have not detected any strong and brief increase of the radio flux
density such as the intense radio flares of the galactic superluminal
sources GRS~1915+105 (Rodr\' \i guez et al. 1995, \cite{poo97,fen99a})
or GRO~J1655-40 (\cite{hje95a}, Hannikainen et al. 1999).  It is
possible that such flaring activity might have not been detected due
to the irregular sampling of our observations, as these flares could be
as short as a few days (e.g. GS~2023+338, \cite{han92}). But based on
the similarity of \gx\ with \cx\ (see \S\ 4) and on the lack of radio
flaring activity in \cx\ (\cite{poo99,bro99}), it is unlikely that
flaring behaviour occurs in \gx. We point out that the radio behaviour
during the Very High state might be different from that outlined in
this paper.

On several occasions, no radio emission has been detected from \gx.
These periods are usually associated with state transitions.
In 1998 when \gx\ entered the High-Soft state, we observed a quenching
of the radio emission associated with a suppression of the hard X-ray
emission (\cite{fen99b}).
 
\subsection{Short term variability}

For a source as weak as \gx, it is not possible to generate a statistically meaningful
light curve with a time resolution of less than 10 minutes. 
A typical radio light curve during a 12 hour observation
in the Low-Hard state is displayed in Figure \ref{fig_lc1}.  The radio
flux density is almost constant on timescale of hours (flux variations
with amplitudes lower than $\sim$ 25 \% cannot be excluded). No radio
quasi periodic oscillation (QPOs), similar to GRS~1915+105 (Pooley \&
Fender 1997), are detected in the Low-Hard state.

Following the quenching of the radio emission during the 1998
High-Soft state, we note that the return of the radio emission occured
with an unusually optically thin radio spectrum on 1999 February 13.
The radio light curve for this observation (Figure \ref{fig_gx_lc2})
displayed an oscillation of $\sim$ 130 minutes with an amplitude of
$\sim$ 30 \% of the average flux density. It is also possible that the
oscillation at 8640 MHz preceeds the one at 4800 MHz by a few minutes,
but this is limited by the sensitivity of ATCA. The rise time for
radio emission is $\sim$ 100 minutes, which is the fastest variation
detected in the radio regime from \gx. In the 10 seconds lightcurve, there
is no indication that the radio light curve is made of individual spikes. 

Interstellar scintillation as the origin of this oscillation in the radio regime
can be ruled out. Indeed, this effect is known to have a strong wavelength dependence
(Romani et al. 1986), which is not observed here as the radio oscillations 
are essentially identical at both frequencies. We should also note that the 
``oscillations'' in \grs\ have also been seen to be correlated with infrared 
and X-ray activity (Pooley \& Fender 1997, Mirabel et al. 1998) and therefore
cannot result from interstellar scintillation.

The X-ray emission (2-100\,keV PCA data) on this date also showed
unusually large variations of approximately greater than 2\% rms variability
on greater than 10 minutes timescales (i.e., frequencies $\le 10^{-3}$ Hz), 
as opposed to the more usual approximately less
than 0.5\% rms found in the Low-Hard state (Nowak et al. 1999).  These
X-ray observations occurred strictly simultaneously with the radio
observations and will be discussed in further detail in a future paper.  In
Figures \ref{fig_osc} and \ref{fig_osc2}, we show the mean-subtracted X-ray and 4800 MHz radio lightcurves
(with the former being multiplied by a factor of 30).  Unfortunately, the
X-ray lightcurve (RXTE) shows a number of gaps due to passage through the South
Atlantic Anamoly.  However, the first X-ray peak, in terms of duration and
relative amplitude, is consistent with preceding the final radio peak by
approximately 10,000 seconds.  Longer simultaneous and uninterupted
X-ray/radio observations, however, are required to confirm if such
behaviour is characteristic of such a Soft to Hard X-ray transition state.

Based on the interpretation of the radio oscillations in \grs\
(\cite{poo97, bel97, mir98}), it is possible that these oscillations
correspond to discrete ejection events associated with state
transitions.

In \grs, the radio oscillations rise when the X-ray dips (Pooley \& Fender 1997, Mirabel et al. 1998).
We might be seeing directly (Figure 5) the same phenomena without the need to make the shift for Figure 6.
If this was the case, the PCA data should show a significant hardening during the dip, as the inner 
accretion disk region disappears.

\section{Characteristics of the radio emission}

\begin{figure}
       \resizebox{\hsize}{!}{\includegraphics{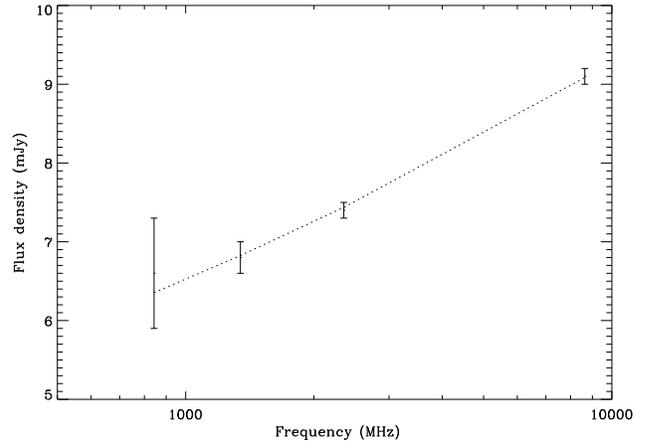}}
        \caption{Typical radio spectrum of \gx\ in the Low-Hard state) from 843 MHz to 9024 MHz, taken around MJD 50484. Also
        shown the best fit is a single power law of spectral index +0.15 $\pm$ 0.04}
        \label{fig_spectre}
\end{figure}

ATCA observations have been performed simultaneously at two
frequencies, giving an estimate of the radio spectrum for every
observation. The spectral indices (using a fitting procedure if more
than two measurements are available, or $\alpha$ = $\Delta$
log(S$_\nu$) / $\Delta$ log($\nu$), i.e. S$_\nu$ $\propto$
$\nu^\alpha$ ) are listed in Table \ref{tab_radio}. The spectrum is usually flat or
slightly inverted with spectral index of $\sim +0.1$.  This inverted spectrum
extends from 843 MHz to 9024 MHz (Figure \ref{fig_spectre}). On a few
occasions, there are some strong indications that this inverted synchrotron spectrum
extends to the near-infrared (\cite{cor00}).
A persistent inverted radio spectrum is not common for most radio emitting
X-ray binaries (but see discussion below on \cx).

Apart from the possible hour ($\sim$ 100 minutes) variability, the
radio flux density of \gx\ is variable on timescales of a day. The
size of the emitting region (see Table \ref{tab_time}) cannot be larger than the minimum
variability time scale times the light velocity (assuming no Doppler boosting).  This implies an
upper limit to the size of the radio emitting region of $\sim$ 2.6
$\times$ 10$^{15}$ cm (or $\sim$ 170 a.u.). A distance of 4 kpc (\cite{zdz98,cor99}) implies an angular
diameter of the source smaller than $\sim$ 43 mas.
This upper limit is reduced to 1.8 $\times$ 10$^{14}$ cm (i.e. 12 a.u.) if we take into account
the radio rise time of $\sim$ 100 minutes on 1999, February 13, which is equivalent to a maximum
angular size of 3 mas.

We can estimate a lower limit to the brightness temperature, T$_B$,
according to the relation: S$_\nu$ = $\Omega$ B$_\nu$(T$_B$). $\Omega$
= $\pi\ \theta^2$/4 is the solid angle subtended by a spherical
emitting region of average angular diameter, $\theta$, and B$_\nu$ is
the Planck function.

\begin{table}[h]
\caption{Constraints on the brightness temperature (T$_B$) and the size of the radio emitting region
from the minimum variability time scale of the radio flux density.}
\label{tab_time}
\begin{center}
\begin{tabular}{lcccc}
\hline
\hline
\noalign{\smallskip}
Variability        &  \multicolumn {3} {c} {Size}                &  T$_B$           \\
\cline{2-4}
\noalign{\smallskip}
		   &   (mas)   &   (cm)               &   (a.u.) &  (K)             \\
\noalign{\smallskip}
\hline
\noalign{\smallskip}
One day 	   &   $<$ 43  &  $<$ 2.6 \xd$^{15}$  & $<$ 170  &  $>$ 1.3 \xd$^7$ \\
100 minutes        &   $<$ 12  &  $<$ 1.8 \xd$^{14}$  & $<$ 12   &  $>$ 2.6 \xd$^7$ \\
\hline
\end{tabular}
\end{center}
\end{table}

From this, we find that the brightness temperature is greater than 1.3
$\times$ 10$^{7}$ K (or 2.6 $\times$ 10$^{7}$ K for the oscillations
at 4800 MHz).  A coherent emission mechanism implies a smaller size
for the emitting region and a higher brightness temperature.  The high
brightness temperature, associated with the inverted frequency
spectra, suggest that \gx\ emits via self-absorbed (or optically
thick) non-thermal synchrotron radiation from relativistic electrons.

A lower limit on the size of the emitting region (see Table \ref{tab_ic}), obtained from the
inverse Compton losses limit (T$_B$ $\la$ 10$^{12}$ K) (see Wilms et
al. 1999), gives a minimum size of 0.07 mas at 843 MHz, i.e. 4.8 $\times$ 
10$^{12}$ cm or 0.4 a.u. (higher frequencies could arise from a smaller region).

\begin{table}[h]
\caption{Inverse Compton losses limit (see text) and the size of the radio emitting region at 843 MHz.}
\label{tab_ic}
\begin{center}
\begin{tabular}{lccc}
\hline
\hline
\noalign{\smallskip}
T$_b$                  &  \multicolumn {3} {c} {Size}                   \\
\cline{2-4}
\noalign{\smallskip}
    (K)                &   (mas)   &   (cm)           &   (a.u.)        \\
\noalign{\smallskip}
\hline
\noalign{\smallskip}
T$_B$ $\la$ 10$^{12}$  &   $>$ 0.07   &  $>$ 4.8 \xd$^{14}$  & $>$ 0.4  \\
\hline
\end{tabular}
\end{center}
\end{table}

We note that if for the observation of 1999 February 13 the 10\,ks
X-ray/radio delay is real, then this implies a minimum propagation speed of
$10^{-3} c$ from the X-ray emitting material to the radio emitting
material.

The interesting characteristics of the radio spectra of \gx\
during the Low-Hard state are that they are
continuously flat or inverted for a radio source at about the same
intensity level. This is very similar to the radio
emission of Cyg~X--1 in its Low-Hard state (\cite{mar96,poo99},
Brocksopp et al. 1999) or the Z sources (e.g. Hjellming et al. 1990a,
Hjellming et al. 1990b). The radio spectrum of \cx\ is flat (i.e. $|\alpha | \leq 0.15~(3\sigma)$  ) up to 220
GHz (\cite{fen00b}) with an average flux density of $\sim $ 14 mJy.
Although similar, the radio spectrum of \cx\ is significantly 
much flatter than the one from \gx.
For most radio emitting X-ray binaries, optically thick synchrotron
emission is usually associated with the beginning of a radio outburst,
which then evolves to an optically thin decaying emission
(\cite{han92,rod95}, Fender et al. 1999a, Kuulkers et al. 1999),
believed to be the results of the ejection and expansion of
relativistic plasma, consistent with the overall picture of the van der
Laan model (\cite{vdl66}).

As noted by Hjellming and Han (1995), stable emission from a radio
source is difficult to maintain as relativistic plasma tends to expand
and therefore implies a decaying radio emission. Relativistic plasmas
are almost impossible to confine, therefore one needs a continuous
injection of plasma in the radio emitting region in order to produce
the stable radio emission. The optically thick spectra are produced
from an inhomogeneous source with a range of optical depths. Therefore,
the radio emission of \gx\ could be explained if one invokes
continuous injection of relativistic particles at the base of a
compact and conical jet [e.g. Blandford \& K$\ddot{\mathrm{o}}$nigl
(1979) or Hjellming \& Johnson (1988)].  In this model, the higher
radio frequencies would come from a region closer to the base of the
jet (where the optical depth is too high for the lower frequencies).

Pooley et al. (1999) have reported modulation of the radio emission
from Cyg~X--1 at the orbital period of 5.6 days, with a stronger
modulation at higher frequency. This indicates that the higher
frequencies come from a region closer to the compact object and
therefore at the base of the jet, which is in agreement with the
conical jet model, such as the one of Hjellming \& Johnson (1988) and
also in agreement with basic Blandford and K$\ddot{\mathrm{o}}$nigl
jet model.

It is interesting to note that a compact and continuous jet has now
been detected in Cyg~X-1 (\cite{sti98,sti00}) from VLBA observations.
The radio emitting region of \cx\ is 2 $\times$ 6 mas with the
smallest synthesised beam at 9 GHz (Fender 1999, private
communication), which is equivalent to a typical size of $\sim$ ten
astronomical units. This is about a factor 10 smaller than the physical size of
the compact jet reported by Mirabel \& Rodr\' \i guez (1999) for \grs\
in a weak optically thick radio state (similar to the Low-Hard state
of \gx\ and \cx).

Such inverted or flat radio spectra, associated with weak radio
emission, are commonly found in the compact core of Active Galactic
Nuclei and are usually associated with small scale (or compact) jets
(\cite{mar85,fal96}).  But as noted in Fender et al. (2000b), the
spectra of compact jets in X-ray binaries are much flatter than ``flat
spectrum'' AGNs where a high frequency cut-off is observed in the
millimeter regime.  Based on these similarities with Cyg~X--1 and AGN,
it is very likely that the radio emission from \gx\ arises from a
compact, continuous and conical jet. This is consistent with the
results from Wilms et al. (1999), which showed that radio emission
from \gx\ originates from regions larger than the whole X-ray binary
system.  The above derived limits on the size of the emitting region
in \gx\ are consistent with the size of the radio emitting region of
\cx. The spectral index of the radio spectra might be related to the 
inclination  of the compact jet, as suggested in Falcke \& Biermann (1999).

\section{Linear polarisation}

During a session of three full 12 hours observations with ATCA in
February 1997, \gx\ was bright enough in radio and the observation
long enough that we were able to detect a significant amount of linear
polarisation from the compact core of \gx. For each observation (1997
February 4, 11 and 18) a level of $\sim$ 2\% linearly polarized
emission at 8704 MHz is observed with a position angle of the electric
field vector of --58.7 $\pm$ 8.3\degr, --74.4 $\pm$ 4.7\degr\ and
--60.0 $\pm$ 6.3\degr\ respectively, i.e. a nearly constant direction.
This is the level expected from optically thick synchrotron emission
if one takes into account cellular depolarisation (the theoretical
level should be around 12 \%, Longair 1994).  Measurements of the total intensity at
two frequencies indicate that the radio spectra of \gx\ were optically
thick, which is usual in the Low-Hard state.

Linear polarisation was again detected on 1999 March 3 (two years
later) at a higher level of $\sim$ 5 \% (both at 4800 MHz and 8640
MHz) during an optically thin radio event associated with the return
of radio emission after the 1998 High-Soft state. The position angle
of the electric field vector was --65.6 $\pm$ 5.3\degr\ at 8640 MHz, in
agreement with the orientation observed two years before.  
This suggests that the compact jet component dominates the linearly polarised radio emission.
It is important to note that this constant angle indicates a favoured axis
in the system, possibly related to the compact jet.   
The extended feature reported by Fender et al. (1997a) has a direction
consistent with this favoured axis. 
For optically thick synchrotron emission, the electric
field vector of the emitted radiation is parallel rather than
perpendicular to the magnetic field (\cite{lon94}). Its direction is
indicated by the solid lines in Figure \ref{fig_atcafeb97}.  This
implies that the magnetic field is quite ordered within the compact
jet and that its averaged orientation is relatively stable over a two
years period.

We note a polarisation angle of --45.4 $\pm$ 5.9\degr\ at 4800 MHz on
1999 March 3, possibly indicative of Faraday rotation. From the minimum
apparent rotation of $\sim$ 10\degr\ between 4800 and 8640 MHz, we can get 
a minimum rotation measure of $\sim$ 40 rad m$^2$, which is quite large.
Detection of linear polarisation in \gx\
confirms that the mechanism at the origin of the radio emission is
synchrotron and \gx\ is the second radio emitting X-ray binary, after
GS~2023+338 (\cite{han92}), to show linear polarisation during an
optically thick radio emission state.

Recently, Fender et al. (2000a) reported the discovery of circularly polarised radio emission
in SS~433. This is the first such detection in a radio emitting X-ray binary.
Concerning \gx, no circular polarisation has been detected at any time with the best 3 $\sigma$
upper limit of 0.7\% at 8640 MHz. This is still above the detection limits in SS~433 (0.6\% at 1400 MHz
decreasing to 0.3\% at 8640 MHz).

\section{Imaging the persistent radio jet in \gx\ ?}

Radio emission from \gx\ involves a compact radio jet, therefore it is
of prime interest to search for this jet in our data. But detection of
such a compact jet is beyond the scope of ATCA observations, which can
achieve spatial resolution of only $\sim$ 1 arcsec. ATCA observation
can be useful to search for large scale structure, similar to the
double-sided jet of \1e\ (\cite{mir92}) or GRS~1758--258 (\cite{rod92}).

\begin{figure}
        \psfig{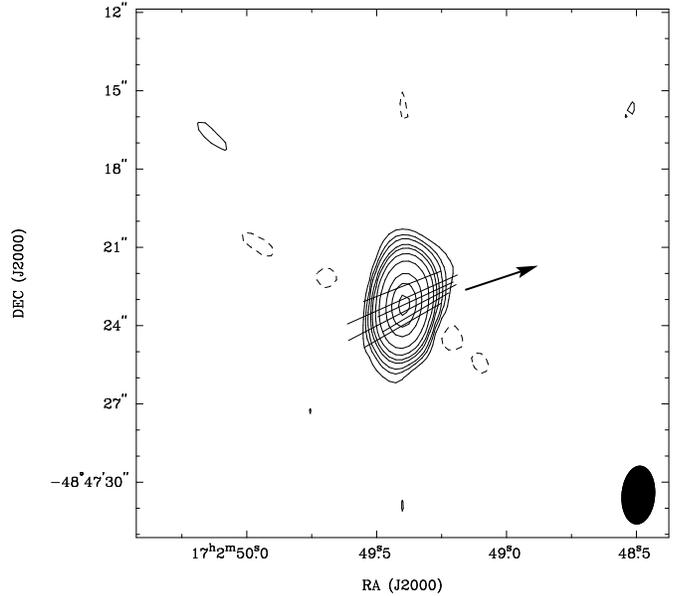}
        \caption{\gx\ at 3.5 cm in February 1997. The peak total intensity is 8.7 mJy beam $^{-1}$, the contour interval (CI, 
taken as the rms in the final image) is 
4 $\times$ 10$^{-2}$ mJy beam $^{-1}$, and contours at --3, 3, 6, 9, 15, 20, 30, 50, 100, 150 and 200 times the CI 
are plotted. A shaded ellipse in lower left corner shows the restoring beam (FWHM). Positive contours are 
solid lines, and negative contours are dashed lines. The orientation of the lines shows the position angle of the 
electric field vector of the linearly polarized signal, and the lengths of the lines are in proportion to the 
strength of the linearly polarized signal. The arrows indicate the approximate position angle of the possible jet reported by
Fender et al. (1997a).}
        \label{fig_atcafeb97}
\end{figure}

The most sensitive observations, coupled with the best available
resolution, have been performed in February 1997. They consist of
three full 12 hours observing runs spaced by a week with the 6A array,
giving baselines ranging from 0.34 km to 6 km (these observations have
been partly presented in \cite{cor97}).  As ATCA is able to observe
simultaneously at two frequencies, we have selected both of them in
the 3 cm band at 8384 MHz and 9024 MHz. We have been using the
multi-frequency synthesis technique to improve our spatial resolution
(\cite{sau94}), i.e. all 32 spectral channels of the 128 MHz bandwidths
of each frequency have been retained at their original position and
have been simultaneously used during the imaging process. The two
frequencies have been chosen in order to complement each other and
provide the best u - v coverage in the 3 cm band. In order to reduce
any artifacts due to phase errors, we have used a cycle of 8 minutes on
\gx\ and 2 minutes on the closest phase calibrators B1646--50 ($\sim$
4 degrees away from \gx).
The size of the synthesised beam is typically 1.3\arcsec $\times$
2.3\arcsec.  A natural weighting map (providing the best signal to
noise ratio) is displayed in Figure \ref{fig_atcafeb97} (this image
combines the three ATCA runs).

The radio map of \gx\ is compatible with a single point source,
without any extended emission around the core, ruling out the presence
of a persistent large scale radio jet.  A uniform weighting map of \gx, which gives a higher resolution
(up to 0.6\arcsec\ E-W) but higher rms noise, is also compatible with a point source.
If a jet exists in \gx, it has
to be very compact with angular size of less than 1\arcsec, i.e. lower
than 0.02 pc or 4000 AU for a distance to the source of 4 kpc. This is
consistent with the size of the compact jet recently detected in
\cx. \gx\ is not bright enough for standard VLBI observations from the
southern hemisphere at present.

We note that these ATCA observations have been performed when \gx\ was
in a Low-Hard state.  A possible large scale jet has been reported by
Fender et al. (1997a), also during a Low-Hard state, but our
observations failed to confirm it. Their report of a jet-like feature
had been preceded by a period of quenched radio emission (see \S\
\ref{tjd10150}). It is also possible that Fender et al. (1997a)
observed a transient phenomenon.

We have also generated a radio image of \gx, combining a wide range of
array configurations (from the 375 m to the 6 km array). This shows
that there is no synchrotron nebula around the point radio source,
such as the one detected around Cir~X--1 (\cite{ste93}). The best
position of the radio counterpart is: $\alpha$(2000) =
17$^h$02$^m$49.4$^s$ and $\delta$(J2000) = --48\degr 47\arcmin
23.3\arcsec\ with a total uncertainty of 0.13\arcsec.

\section{Radio, soft and hard X-rays behaviour}

In 1998, when \gx\ entered the High-Soft state, we observed the
quenching of the compact radio jet (Fender et al. 1999b).  The
transitions from the Low-Hard state to the High-Soft state, and also
the return to the Low-Hard state, have been both accompanied by
an unusually optically-thin synchrotron event lasting around three
weeks. This could possibly be interpreted as discrete ejection events
associated with state transitions, and it could be a general
property of state transitions in black hole binaries.  Following the
results acquired during the 1998 High-Soft state, we now concentrate
on all available data in order to show that the radio:X-ray behaviour
of \gx\ follows a clear pattern.  We displayed in a chronological
order the radio data together with the hard and soft X-ray data (if
available) to highlight the behaviour of \gx\ at these wavelengths.

\begin{figure*}
\resizebox{\hsize}{11cm} {\includegraphics{9629.f9}}
\caption{ Radio, hard and soft X-rays light curves of \gx\ for the period MJD 48350-48750. 
The radio measurements of \gx\ are tabulated in Table \ref{tab_radio}. The hard X-ray measurements (20-100 keV) are those obtained by \batse\ 
and the soft X-ray data (1-20 keV) are from the All Sky Monitor onboard {\em Ginga} (\cite{kit93}).}
\label{fig_a}
\resizebox{\hsize}{11cm} {\includegraphics{9629.f10}}
\caption{Radio and hard X-ray light curves of \gx\ for the period MJD 49200-49550. The radio measurements are from Hannikainen 
et al. (1998).  The hard X-ray measurements (20-100 keV) are those obtained by \batse.}
  \label{fig_b}
\end{figure*}

\begin{figure*}
\resizebox{\hsize}{11cm} {\includegraphics{9629.f11}}
\caption{Same as Figure \ref{fig_b}, but for the period MJD 49550-49750. The ATCA radio measurements of \gx\ are tabulated 
in Table \ref{tab_radio}. }
\label{fig_c}
\resizebox{\hsize}{11cm} {\includegraphics{9629.f12}}
\caption{Same as Figure \ref{fig_b}, but for the period MJD 49750-50050.}
\label{fig_d}
\end{figure*}

\begin{figure*}
\resizebox{\hsize}{11cm} {\includegraphics{9629.f13}}
\caption{Same as Figure \ref{fig_a}, but for the period MJD 50150-50350. The soft X-ray data (1.3-12.2 keV) are from the 
All Sky Monitor onboard {\em RXTE}.}
\label{fig_e}
\resizebox{\hsize}{11cm} {\includegraphics{9629.f14}}
\caption{Same as Figure \ref{fig_e}, but for the period MJD 50350-50600.}
\label{fig_f}
\end{figure*}

\subsection{MJD 48350-48750: 1991-04-04 to 1992-05-08 }

This period corresponds to the first hard X-ray outburst (bright
Low-Hard state) detected by BATSE (Harmon et al. 1994, \cite{rub98}).
In addition to the BATSE light curve in Figure \ref{fig_a}, we have
also displayed the soft X-ray (1-20 keV) light curve from the All Sky
Monitor onboard {\em Ginga} (\cite{kit93}) and the first two
measurements and detection of \gx.  Unfortunately, no {\em Ginga} data
are available after MJD 48531. From the spectral index, we can infer
that these radio observations correspond to optically thin events.
The radio spectrum is becoming flatter ($\alpha \rightarrow 0$) between the two
observations (from a spectral index of --0.64 $\pm$ 0.28 to --0.17
$\pm$ 0.17 ), which is identitical to the end of the 1998 High-Soft
state (\cite{fen99b}).

This hard X-ray outburst started around MJD 48440, the soft and hard
X-rays are correlated until the hard X-rays reached their maximum on MJD
48523. This was followed by a fast decline in hard X-rays when the
source became very soft with a steeper power law spectrum (Harmon et
al. 1994), indicative of a transition to a High-Soft state.  This
optically thin radio event was detected (December 1991) about 70 days
after the beginning of the High-Soft state. According to our results
on the 1998 High-Soft state of \gx\ (\cite{fen99b}), the radio event
at MJD 48590 probably corresponds to a discrete ejection event (or the
refuelling of the compact jet) at the end of the High-Soft state,
which was therefore very short.  There is a corresponding slight
increase in the hard X-ray flux (from \batse) at the same time ($\sim$
MJD 48590).

The {\em OSSE} spectrum in November 1991 (Grabelsky et al. 1995) is
consistent with \gx\ still being in its High-Soft state. Trudolyubov
et al. (1998) reported that \gx\ was in the Off state on
February-March 1992 (one $\sigma$ upper limit of 19 mCrab (35-150 keV)
on 1992 March 7), in agreement with the previous picture and
indicative of a very short High-Soft state.
It should be noted that the return from the 1998 High-Soft 
state to the standard Low-Hard state has also been followed by
a transition to an X-ray Off state (see \S\ \ref{section_off}).

\subsection{MJD 49200-49550: 1993-08-01 to 1994-07-17}

During this period (Figure \ref{fig_b}), a third outburst (these hard
X-ray outbursts are still the Low-Hard state, and should not be
confused with the High-Soft state) was detected by BATSE (Rubin et
al. 1998). The outburst was very similar to the 1991 hard X-ray
outburst, unfortunately no soft X-ray data are available for this
period. The four MOST data points around MJD 49500 do not fit in the
standard radio - hard X rays (see \S\ \ref{section_correl})
correlation scheme (as previously noted by Hannikainen et al. 1998).
Based on the fast drop of the hard X-ray emission at MJD 49425, we can
speculate that it corresponds to a transition to a High-Soft state and
that similarly to December 1991, these MOST data points would
correponds to an optically thin decaying discrete ejection event at
the end of the High-Soft state.  As MOST is operated at one frequency
only, we cannot confirm the optically thin nature of this radio
event. As in  1991, the duration of the High-Soft state would
have been around 70 days.

\subsection{MJD 49550-49750: 1994-07-17 to 1995-02-02 }

Observations during a hard X-ray outburst [B4 in Rubin et al. (1998)]
revealed that the radio emission follow the general trend of the hard
X-ray flux (Figure \ref{fig_c}).  No flaring activity, in the radio
domain, accompanies this outburst. The highest radio flux (12 mJy) is
reported around the maximum of the outburst with a flat radio spectrum
between 4800 and 8640 MHz.

\subsection{MJD 49750-50050: 1995-02-02 to 1995-11-29}

\gx\ is still in its typical Low-Hard state with weaker outbursts
(Figure \ref{fig_d}). Radio emission followed the general trend of
hard X-ray emission. The radio spectrum is again flat.

\subsection{MJD 50150-50350: 1996-03-08 to 1996-09-24} \label{tjd10150}

This is an interesting period (Figure \ref{fig_e}) when radio emission
is not detected several times. And this quenched radio state does not
correspond to a High-Soft state as indicated by the very low flux
detected by RXTE/ASM. The two MOST non-detections around MJD 50200 are
consistent with \gx\ being in the Off state and the following MOST
data points represent a return to a Low-Hard state in agreement with
the optical level of V = 17 mag. on MJD 50215 reported by Smith et
al. (1999b).

A hard X-ray outburst occured after MJD 50250 (1996-06-16). It is
detected in all three regimes (radio, soft and hard X-rays).  The
radio and hard X-ray observations are consistent with a simultaneous
onset, but with a faster rise in radio.  Soft X-rays appeared with a
significant delay of about $\sim$ 15 days but with a faster rise than hard
X-ray and radio emissions.

15 days is significantly longer than the expected viscous timescale at the
outer edge ($\sim~\le$ 400 GM/c$^2$; see Wilms et al. 1999, Zdziarski et al. 1998)
of a spherical X-ray emitting corona, whether it is radiatively efficient
or ``advection dominated''.  Such time delays would be more characteristic of
viscous timescales at the outer edge of the accretion disk. It is well
known that in the Low-Hard state, increased flux tends to imply softer
spectra (see Wilms et al. 1999, and references therein), and therefore we
may be seeing such variations as the disk accretion rate changes on the
viscous timescales of the outer disk. {\em RXTE}, {\em OSSE} and {\em BATSE}
observations during this outburst have been discussed by Smith et al. (1999a)
and Smith et al. (1999b).

At the peak of this outburst (1996 July 12) the radio spectrum was
very inverted with spectral index up to +0.54 $\pm$ 0.05 and it
corresponds to the period when Fender et al. (1997a) reported a
possible jet-like feature in \gx\ on a much larger scale than the
compact jet (a phase error could not be totally ruled out during these
observations).  Prior to this hard-X outburst, radio emission was not
detected from \gx\ during $\sim$ 20 days, with the strongest
measurement giving a 3 $\sigma$ upper limit of 0.6 mJy at 8640
MHz. This is reminiscent of quenched radio emission from Cyg~X--3
prior to major ejection events (\cite{wal96,fen97b}, McCollough et al. 1999).

\subsection{MJD 50350-50600: 1996-09-24 to 1997-06-01}

This period (Figure \ref{fig_f}) illustrates again the overall behaviour of \gx\ in the Low-Hard state, with 
a strong correlation of emission in the radio, soft and hard X-ray domains. The radio spectra are flat, indicative
of the optically thick emission from the compact jet. The observations with the highest signal to noise ratio 
have been  performed in February 1997 ($\sim$ MJD 50490). See Nowak et al. (1999) and Wilms et al. (1999) for
{\em XTE} observations during this period.

\section{A transition to the Off state after the 1998 High-Soft state} \label{section_off}

\begin{figure*}[hbt]
\centering
\psfig{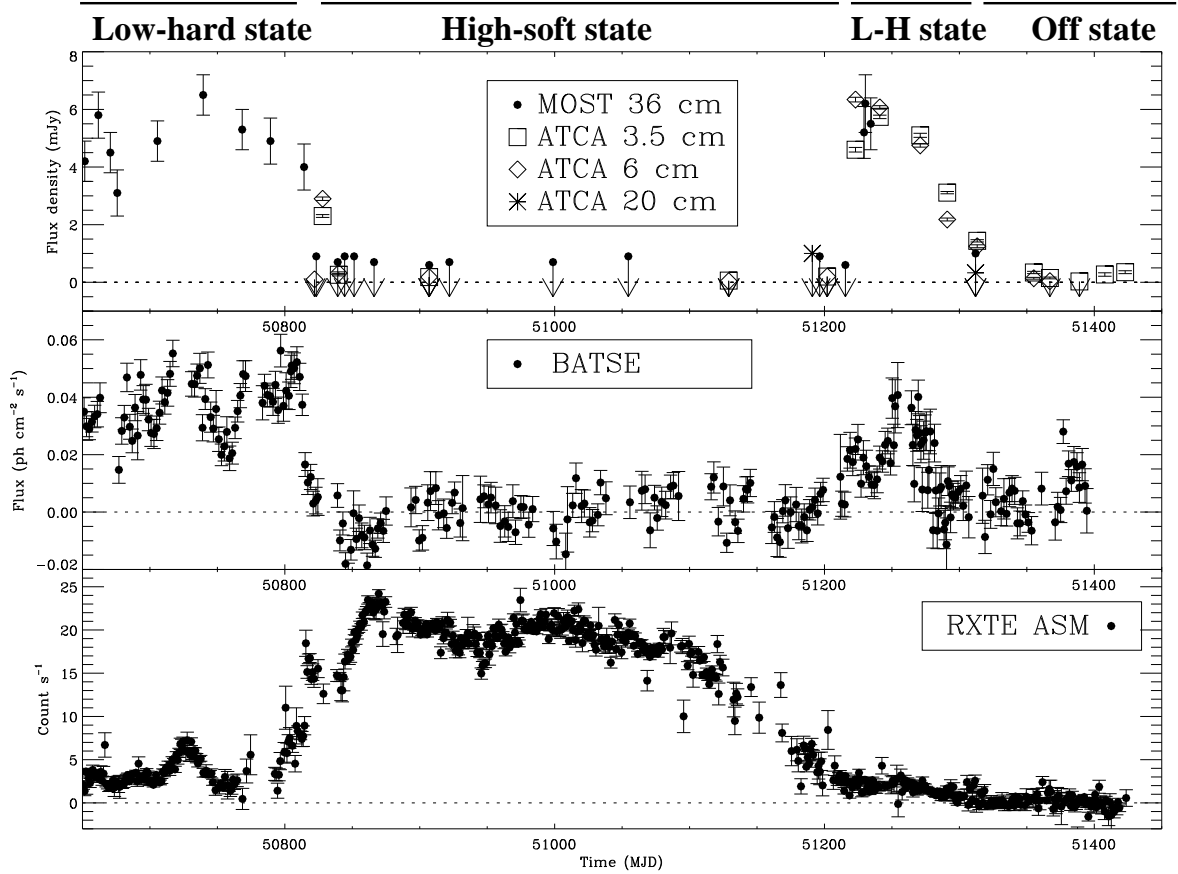}
\caption{Same as Figure \ref{fig_e}, but for the period MJD 50650-51450.}
\label{fig_g}
\end{figure*}

This period (MJD 50650-51450: 1997-07-21 to 1999-09-29) is displayed in
Figure \ref{fig_g}. The first part of this plot has been discussed in
detail in Fender et al. (1999b). It corresponds to the 1998
transition from the Low-Hard state to the High-Soft state
(\cite{bel99}) and then back to the Low-Hard state.  One observation
has been added in Figure \ref{fig_g} (\gx\ was not detected) at the
end of the 1998 soft X-ray outburst on MJD 51215. The important point
to note is that radio emission from \gx\ rose from a level of $<$ 0.7
mJy (1 $\sigma$ at 843 MHz) to 6.3 mJy on MJD 51223 (at 4800 MHz),
i.e. the reappearance of radio emission in less than seven days.  Moreover
this state transition was accompanied by an optically thin synchrotron
event, probably related to a discrete ejection events, similarly to
the beginning of the High-Soft state (\cite{fen99b}).

It is interesting to note that the hard X-ray emission reappears
$\sim$ 2 weeks before the radio emission. This can be understood if
one consider the hot corona as the base of the jet. In the High-Soft
state the corona is believed to shrink to a very small size (e.g. Esin
et al. 1998). 
The return to the Low-Hard state can be understood as a
``refill'' of the corona, which then can later provide relativistic
electrons to the compact jet.

The return to the Low-Hard state has been followed by a smooth
transition to an Off state with radio and hard X-ray emissions getting
weaker as the soft X-rays vanished. \gx\ was detected on 1999, March 3
at $\sim$ 6 mJy and on 1999 July 7 at 0.15 mJy.  On 1999, July 27 the
radio emission was below the detection limit of our observations (3
$\sigma$ upper limit of 0.09 mJy at 8640 MHz). It was  detected
two weeks later at $\sim$ 0.3 mJy, again preceded by an increase in
hard X-rays.

During this transition to the Off state, the radio spectrum was always
slightly inverted, indicative of optically thick synchtrotron
emission. Therefore the compact radio jet is still a characteristic
feature of \gx\ even in the Off state.  The Off state is related to a
lower mass accretion rate, which probably reduces the electron density
in the accretion disk and in the corona and therefore in the compact
jet. From the point of view of the radio emission at least,
the Off state can be viewed as a weaker luminosity version of the
Low-Hard state. Actually, Kong et al. (2000) also found that the Off 
state was an extension of the Low-Hard state, based on {\em BeppoSAX} observations
of \gx\ in the Off state. All the radio observations performed in 1999 have
been simultaneous with soft X-ray ({\em RXTE} and {\em ASCA}) and
optical observations. This will be discussed in more detail in a
forthcoming paper.

\section{Radio, soft and hard X-ray correlations} \label{section_correl}

\begin{figure*}
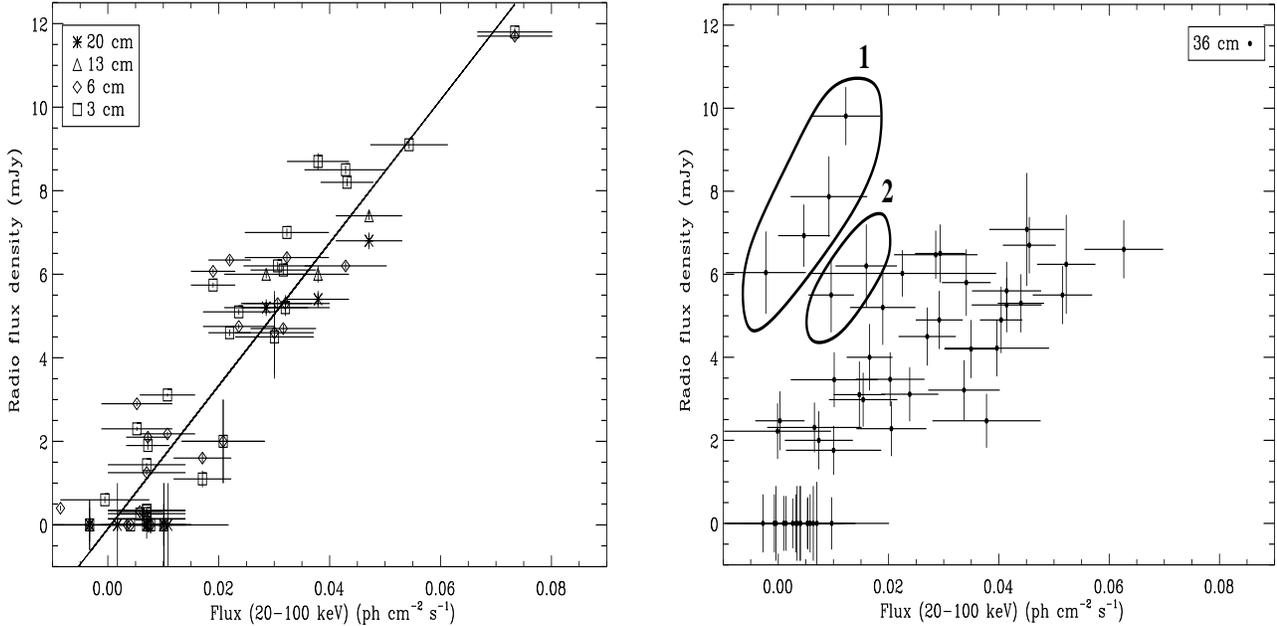

\vspace{0cm}
\hbox{\hspace{0cm}\psfig{figure=9629.f16a,width=8.8cm,height=9cm}\hspace{0cm}
\psfig{figure=9629.f16b,width=8.8cm,height=9cm}}
\caption{Radio flux density from ATCA (left) and MOST (right) observations as a function of hard X-ray flux (20-100 keV),
obtained by \batse. The straight line indicates the linear fit to the measurements.
Events 1 and 2 correspond to possible discrete ejection events associated with X-ray state transtions.}
\label{fig_atca_batse}
\vspace{0cm}
\end{figure*}

\begin{figure*}
\vspace{0cm}
\hbox{\hspace{0cm}\psfig{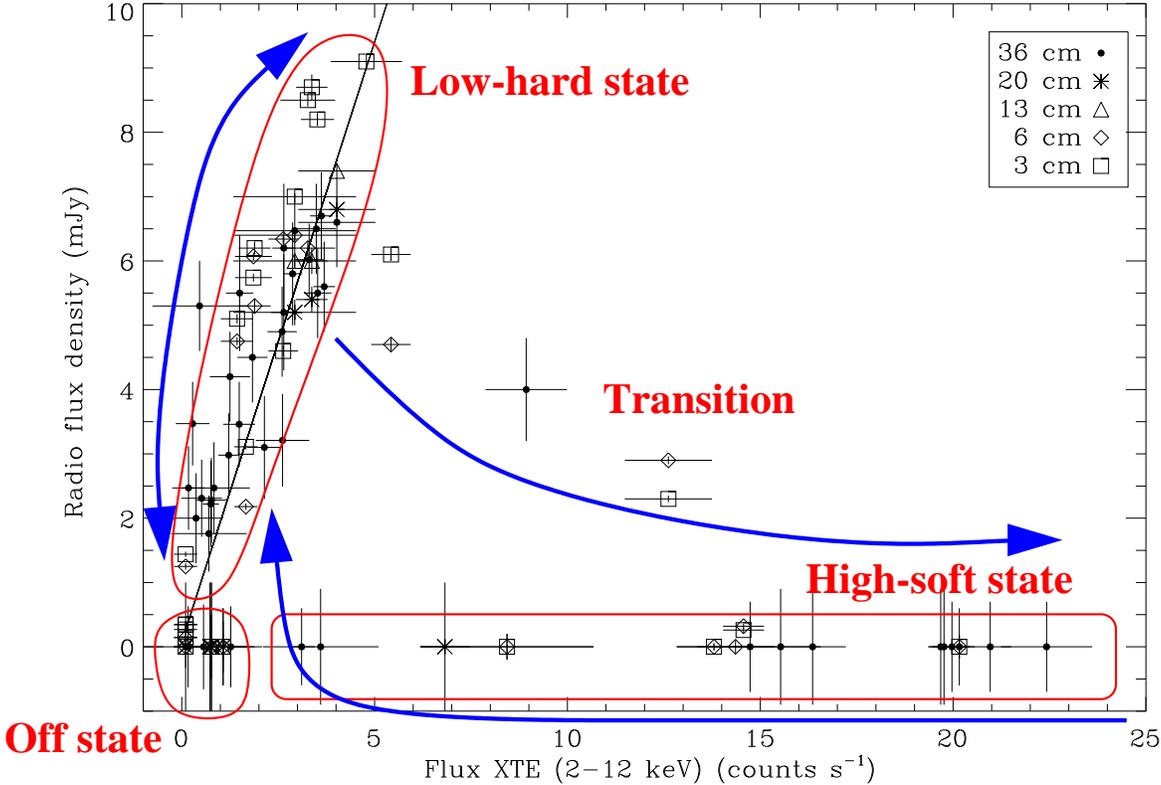}\hspace{0cm}}
\caption{Radio flux density from MOST and ATCA observations as a function of soft X-ray flux. The straight line indicates the
linear fit for the variation of radio flux density with soft X-ray emission in the Low-Hard state. Arrows indicate the
evolution within and between the X-ray states.}
\label{fig_most_batse}
\vspace{0cm}
\end{figure*}

Hannikainen et al. (1998) reported a correlation between the radio
emission at 843 MHz (from MOST observations) and soft and hard X-rays
on timescales of $\sim$ 5 days. In Fender et al. (1999b), we showed
that this correlation doesn't hold during the High-Soft state, and
that radio and hard X-ray emission are both suppressed during the
High-Soft state, following the quenching of the compact jet.

We have used our radio data (spanning over 8 years) to explore this
correlation at several radio frequencies with both ATCA and
MOST. Unlike Hannikainen et al. (1998), we have not performed any
averaging of the measurements with the exception of non simultaneous
radio and X-ray observations.  In Figure \ref{fig_atca_batse}, we plot
the radio flux density (ATCA and MOST) from \gx\ as a function of hard
X-rays (BATSE flux in the 20-100 keV band).  We observe a very good
correlation of radio emission with hard X-ray flux. This correlation
is better seen with ATCA measurements rather than MOST observations,
because of ATCA's higher sensitivity. The correlation is observed at
all radio frequencies (not surprising as the radio spectrum of \gx\ is
nearly flat).  In the figure showing the MOST measurements, two events
(labelled 1 and 2) are clearly separated from the other data
points. As pointed out before, these events are related to possible
discrete plasma ejection events associated with a transition from a
High-Soft state to a Low-Hard state.

The linear Pearson correlation coefficient is 0.93 for the radio and
hard X-ray measurements.  A least-square polynomial fit to the ATCA
data gives the following function between the radio flux density
S$_{\rm cm}$ and the 20-100 keV hard X-ray emission S$_{\rm HXR}$
(in 10$^{-2}$ ph cm$^{-2}$ s$^{-1}$, in the same band the Crab nebula is 
at a level of 0.32 ph cm$^{-2}$ s$^{-1}$):

\begin{equation}
S_{\rm cm} = -0.09 + 1.71~S_{\rm HXR} \phantom{000}
{\rm mJy}
\end{equation}

If we now concentrate on the radio flux density as a function of soft
X-ray emission (2-12 keV RXTE/ASM count rate), the overall picture
seems more complicated at first sight, but this is only because it
displays the various behaviours of \gx\ in its different X-ray states
(Figure \ref{fig_most_batse}). The data points are divided in four
groups, indicative of the observed three different X-ray states and of
the transition between the High and Low states.  As noted before, the
radio emission is quenched in the High-Soft state (which correspond to
high RXTE/ASM count rate and no radio emission).  Althought few measurements have been taken during a
transition from the Low-Hard state to the High-Soft state, they follow
a clear pattern in Figure \ref{fig_most_batse}.

By taking only the radio measurements during the Low-Hard state, we
note that there is also a strong correlation between the radio flux
density and the soft X-ray emission. This correlation extends to very
weak levels of emission (Off state), as the 1999 return to the Low-Hard 
state has been followed by a transition to the Off state with the
radio flux density getting weaker simultaneously with the soft X-ray
emission (Figure \ref{fig_g}).

The linear Pearson correlation coefficient is 0.84 for the radio and
soft X-ray emission in the Low-Hard state.  A least-square polynomial
fit to the ATCA data gives the following function between the radio
flux density S$_{\rm cm}$ and the 2-12 keV soft X-ray emission
S$_{\rm SXR}$ (in count s$^{-1}$, in the same band the Crab nebula is at 
a level of 74 count s$^{-1}$):

\begin{equation}
S_{\rm cm} = 0.07 + 1.87~S_{\rm SXR} \phantom{00} {\rm mJy}
\end{equation}

This analysis shows that there is a very strong 3-way correlation between
the radio flux density, the soft and hard X-ray emission during the
Low-Hard state.
In the Low-Hard state, emission from the inner accretion disk does not contribute
significantly in the soft X-ray band (e.g. Belloni et al. 1999), which emission
is dominated by the Comptonising corona. 
This correlation indicates that there is a strong
coupling between the compact jet (at the origin of radio emission) and
the corona (soft and hard X-rays). A
similar correlation has also been observed in \cx\ during the Low-Hard
state (Brocksopp et al. 1999). As noted in Fender et al. (1999b), it is
possible that the corona is at the base of the compact jet; the high
energy tail of the electron distribution being responsible for the
synchrotron emission observed at radio (and possibly higher) frequencies.

The correlation between the corona and the compact jet may reflect variability
resulting from changes in the accretion rate, therefore implying a 
correlation between the compact jet, the corona and the accretion rate.
The thermal emission from the inner part of the accretion disk is generally not observed 
in the Low-Hard state, but it might also reflect the changes in the
accretion rate. Such questions could be addressed by future observations
with {\em XMM} and {\em Chandra}. We should note that these correlations
may result from the precession of the accretion disk as suggested by
Brocksopp et al. (1999) for \cx.

The fact that the correlation between the radio and the X-ray flux in the Low-Hard
state is consistent with being linear is somewhat difficult to understand
in the context of advection dominated models that postulate large outflows
(e.g., Blandford \& Begelman 1999). In the absence of such large outflows,
standard ADAF models fail to produce, by several orders of magnitude, the
observed radio flux (Wilms et al. 1999).  In a wind-producing ADAF model,
a decrease in luminosity is associated in part, at least, with a further
decrease in radiative efficiency.  This, however, is expected to result in
a greater rate of mass outflow.  Therefore one expects the radio flux to
decrease somewhat more slowly than the X-ray flux, in contrast to the
linear dependence observed.  If in fact the transition to
the Low-Hard state is a transition to an ADAF state, perhaps then the
radiative efficiency remains relatively constant with further decreases in
flux.

Following the equation of Marscher (1983), hard X-rays resulting from
Synchrotron Self-Compton emission from the compact jet of \gx\ can be
neglected as it is negligible compared to the contribution from the
corona.  A similar conclusion has been drawn by Brocksopp et
al. (1999) for \cx.

\section{Conclusions}

The following list summarizes the conclusions we have been able to
draw from this multi-wavelength analysis of the behaviour of \gx:

\begin{itemize}
\item 
{In the Low-Hard X-ray state, we observe a very strong 3-way
correlation between the radio, soft and hard X-rays emission, believed
to be the result of a coupling between the Comptonising
corona and a compact jet.  Models of persistent accreting
black holes would need to take into account this coupling in the
Low-Hard state and the quenching of the radio jet in the High-Soft
state. The Off X-ray state is consistent with being a lower-luminosity
Low-Hard state, in all three energy bands.}
\item
{The radio spectrum is flat or slightly inverted. A similar behaviour
is observed in the black hole \cx.  The radio emission can be
understood as synchrotron emission from a compact jet, with quasi-continuous 
injection of relativistic plasma. }
\item{Transitions from the Low-Hard state to the High-Soft state (or
vice versa) are possibly associated with discrete ejection(s) of
expanding relativistic plasma.}
\item{The pattern of behaviour observed in radio, soft- and
hard-X-rays is sufficient to explain in broad terms all the
observations over several years.}
\item{Linear polarisation has been detected from the radio emission of
\gx\ with a nearly constant polarisation angle, which points to a
favored axis in the system. This direction is almost certainly related
to the direction of the outflow, and in turn to the inner disc and/or
black hole rotation axes.}
\end{itemize}

\begin{acknowledgements}

The Australia Telescope is funded by the Commonwealth of Australia for
operation as a National Facility managed by the CSIRO. The MOST is
owned and operated by the University of Sydney, with support from the
Australian Research Council and the Science Foundation within the
School of Physics.  RXTE ASM results are kindly provided by the
ASM/RXTE teams at MIT and at the RXTE SOF and GOF at NASA's GSFC. We
are grateful to D. Campbell--Wilson and D. Hunstead for their help
with the MOST observations. We would like to thanks Dr R. Ramachandran for useful
discussion on interstellar scintillation and Prof. S. Kitamoto
for provinding the Ginga ASM data. S.C. would like to thank C. Bailyn,
C. Chapuis., C. Gouiff\`es, R. Jain, M. Mouchet, H. Negoro, R. Ogley, B. Sault,
H. Sol, M. Tagger and J. Wilms for useful discussions and A. Harmon,
M. McCollough and C. Robinson for their help with the BATSE data. RPF
was supported during the period of this research by EC Marie Curie
Fellowship ERBFMBICT 972436.

\end{acknowledgements}

\end{document}